\documentclass{article}
\usepackage{float}
\usepackage{graphicx}

\usepackage[numbers]{natbib}
\usepackage[preprint]{nips_2018}

\usepackage[utf8]{inputenc} 
\usepackage[T1]{fontenc}    
\usepackage{hyperref}       
\usepackage{url}            
\usepackage{booktabs}       
\usepackage{amsfonts}       
\usepackage{nicefrac}       
\usepackage{microtype}      
\usepackage{subcaption}
\usepackage{caption}

\title{Learning a Generative Model of Cancer Metastasis}

\author{
  Benjamin Kompa \\
  \texttt{kompa@fas.harvard.edu} 
  \And
  Beau Coker\\
  \texttt{beaucoker@g.harvard.edu}
}

\begin{document}

\maketitle

\begin{abstract}
  We introduce a Unified Disentanglement Network (UFDN) trained on The Cancer Genome Atlas (TCGA). We demonstrate that the UFDN learns a biologically relevant latent space of gene expression data by applying our network to two classification tasks of cancer status and cancer type. Our UFDN specific algorithms perform comparably to random forest methods. The UFDN allows for continuous, partial interpolation between distinct cancer types. Furthermore, we perform an analysis of differentially expressed genes between skin cutaneous melanoma (SKCM) samples and the same samples interpolated into glioblastoma (GBM). We demonstrate that our interpolations learn relevant metagenes that recapitulate known glioblastoma mechanisms and suggest possible starting points for investigations into the metastasis of SKCM into GBM. 
\end{abstract}

\section{Introduction} 
Deep learning is being applied to many difficult problems in genomics and medicine. Alipanahi et al. used deep learning to learn site specific binding patterns of DNA and RNA-binding proteins \cite{Alipanahi2015-as}. Zhou et al. were able to predict non-coding variants using deep learning \cite{Zhou2015-yn}. Google has even produced an improved variant caller known as DeepVariant \cite{Anderson2018-fa}. 

More specifically, deep learning has been applied to understanding cancer prognosis. Chaudhary et al. were able to robustly predict survival in liver cancer \cite{Chaudhary2018-pf}. Cruz-Roa et al. leveraged deep learning to quantify the extent of breast cancer tumors in imaging data \cite{Cruz-Roa2017-oi}. Other groups have trained networks to identify metastatic breast cancer  and lymph node metastasis \cite{Wang2016-eg, Bejnordi2017-zk}.

Nevertheless, there is little work in machine learning being done on what changes are occurring at a gene expression level in cancer samples. Understanding the genomic basis of cancer will yield better treatments and prognosis for patients \cite{Al-Lazikani2012-rn}. There are significant questions remaining in oncology about the relationships between different cancer types. For instance, while there is an association between melanoma, a type of skin cancer, and glioblastoma, a type of brain cancer, little is known about the molecular underpinnings of this relationship \cite{Desai2008-cj, Scarbrough2014-wj}. 

Recently, deep generative models such as variational auto encoders (VAEs) and generative adversarial networks (GANs) have made large advances in image, audio, and text generation \cite{Hsu2017-ov, Larsen2015-fz, Pu2016-ln}. VAEs and GANs learn generative distributions on lower-dimensional encodings of input data \cite{Way2018-yi}. VAEs have found genomic applications. Rampasek et al. applied VAEs to learn drug responses based on gene expression data \cite{Rampasek2017-kv}. Way et al. trained a VAE called Tybalt to encode The Cancer Genome Atlas (TCGA) \cite{Way2018-yi}. Huang et al. have developed a theory of cancer development as a progression along a low dimensional space, justifying exploration of cancer metastasis using machine learning algorithms that learn low dimensional representations \cite{Huang2009-wr}. 

A new VAE-GAN hybrid architecture known as the Unified Feature Disentanglement Network (UFDN) learns fundamental features that distinguish input domains \cite{Liu2018-jg}. For multiple input data types, such as photographs, sketches, and watercolor paintings, the UFDN learns  an VAE encoding of the data domains and trains a discriminator in the latent space to discriminate between domain types. Then, the UFDN can subsequently encode data from one domain and decode the data into a different domain\cite{Liu2018-jg}. An additional GAN distinguishes between real/fake images in the pixel space to promote high quality decodings\cite{Liu2018-jg}. 

In this work, we apply this new UFDN architecture to TCGA RNA-Seq data and learn a latent space embedding that allows us to convert between different cancer types given gene expression data. Given gene expression levels in a cancer sample of domain $A$, we can predict gene expression levels as if that cancer sample were of domain $B$. This represents a generative, personalized model of metastasis. We can sample points in our latent space encoding and decode them into any new cancer domain. 

Additionally, we can partially interpolate between cancer domains. UFDN decoding is not strictly binary---input data can be decoded into a mix of output domains. We investigate \textit{partial interpolations} of one cancer type into another, mimicking the progressive nature of metastasis. 

We analyze the performance of our TCGA-trained UFDN on two tasks: predicting whether a sample is from cancerous or normal tissue and predicting which cancer sub-type a sample consists of. Additionally, we investigate partial interpolations from skin cutaneous melanoma (SKCM) TCGA samples to glioblastoma (GBM) by looking at differential expression of genes. We compute metagenes that summarize gene expression changes using integrative non-negative matrix factorization. Finally, we analyze Gene Ontology (GO) term enrichment in highly activated metagenes for each interpolated dataset.

\begin{figure}[htbp]
    \centering
    \includegraphics[width=\textwidth]{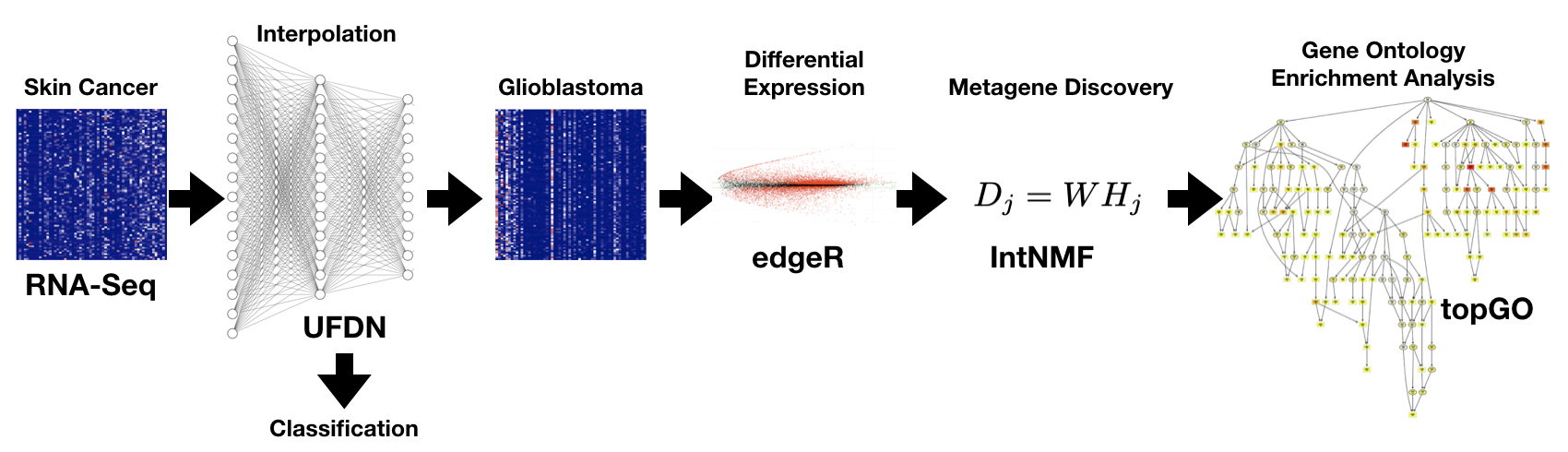}
    \caption{The overall workflow of our project. We aimed to identify the crucial changes in gene expression as cancer metastasizes from the original location to a new location. We encoded RNA-Seq samples from skin cutaneous melanoma, decoded them into glioblastoma, and then applied 3 bioinformatics tools to analyze which sets of genes were changing between cancer types.}
    \label{fig:figure1}
\end{figure}

\section{Methods}

\subsection{Data Preprocessing}
The data consisted of 10,433 samples of RNA-Seq gene expression levels across 33 cancer types for 20,501 genes from TCGA obtained via the R Package \texttt{curatedTCGA} \cite{Cancer_Genome_Atlas_Research_Network2013-pt, Ramos2018-cx}. For the purpose of this work, we only considered the RSEM (RNA-Seq by Expectation Maximization \cite{Li2011-ra}) normalized expression levels. We divided 70\%, 20\%, and 10\% to train, test, and holdout datasets, respectively.

Tybalt demonstrated that preprocessing gene expression levels by scaling gene-wise expression levels (across all samples) to between 0 and 1 yields a trainable latent space \cite{Way2018-yi}. We adapted this procedure by first clipping
expression levels to fall within 3 standard deviations from the mean of gene-wise expression levels followed by the same min-max normalization of Tybalt \cite{Way2018-yi}. 
\subsection{UFDN}
\subsubsection{Theory}
Liu et al. develop a UFDN as a combination of an encoder $E$, a generator $G$, and two discriminators: $D_v$ in the latent space and $D_x$ in the pixel space \cite{Liu2018-jg}. In our application, pixel space is replaced by ``gene expression space.'' $E$ takes input data and encodes it in a latent space. In our UFDN, we encode gene expression using fully connected networks. $D_v$ learns to discriminate between domains, or cancer types. Then generator $G$ uses a latent space encoding and a domain vector $d_v$ to produce gene expression data in domain $v$ \cite{Liu2018-jg}. Our UFDN uses $d_v\in\mathbb{R}^{33}$ since there are 33 cancer types in TCGA.

We define a \textit{partial interpolation} with parameter $p\in[0,1]$ of an input of domain $c$ to domain $\hat{c}$ to be the decoding of the input into into a composition of domains $c$ and $\hat{c}$, with weight $p$ given to domain $\hat{c}$. That is, the domain vector of the partial interpolation has components $d_{v_{\hat{c}}}=p$, $d_{v_c}=1-p$, and remaining components zero. For instance, a 0.25-GBM interpolation means an input has been decoded with $d_{v_{GBM}}=0.25$ and original domain entry is $0.75$. 

In the pixel space (or gene expression space), $D_x$ learns to distinguish between samples that have been decoded to their original domain $c$ or a new domain $\hat{c}$ \cite{Liu2018-jg}. The network is trained by iterative stochastic gradient updates to $E$, $D_v$, and $D_x$. For a more detailed exposition of the architecture of and gradient updates for training the UFDN, please see Section 3 of Liu et al. 2018 \cite{Liu2018-jg}. 

The encoder $E$ and generator $G$ are single layer networks, each with 500 hidden units, that learn a 100 dimensional latent space. The feature space discriminator $D_v$ is a single layer network with 64 hidden units and the pixel space discriminator $D_x$ is a two layer network with 500 and 100 hidden units. All networks are fully connected with leaky relu activation functions. We use 50,000 iterations of Adam updates with a learning rate of $10^{-4}$.

\subsubsection{Classification Tasks}
We attempted two classification tasks using the UFDN. The first was classifying a sample as tumor or normal. This is referred to as the cancer status task. The second task was predicting cancer domain, one of 33 sub-types in the TCGA. 

In order to solve these tasks, we developed 3 algorithms using UFDN: 
\begin{itemize}
    \item UDFN-MSE: classify a sample's type by encoding the sample and decoding it into all 33 domains, predicting the type of the domain with lowest reconstruction error as defined by mean square error (MSE). 
    \item Unsupervised UFDN: Inspired by the unsupervised domain adaptation experiments from Liu et al.\cite{Liu2018-jg}, this algorithm predicts cancer status by encoding a sample into the latent space, then decoding it into the mesothelioma domain, regardless of input domain. We trained a random forest classifier to predict cancer status on mesothelioma training data. Use the prediction of this classifier to predict cancer status in the original input domain. The motivation for this approach is that the classifier trained on mesothelioma data is strong but the test data of interest is of a different cancer type.
    \item Semi-supervised UFDN: A hybrid of the two above algorithms used to predict cancer status and type. First, predict cancer type using UDFN-MSE. Then, predict cancer status using a random forest classifier trained on that specific type's status data. 
\end{itemize}

\subsection{Interpolation Analysis} 
We encoded 95 samples of SKCM (skin cutaneous melanoma) from our test set partition of the TCGA into our latent space using our trained UFDN. Then, we interpolated the samples into glioblastoma (GBM) at four different fractions of interpolation: 25\%, 50\%, and 75\%, and 100\%. The 100\% interpolation represents a prediction of gene expression levels of the SKCM samples as GBM per sample. 

In order to analyze how gene expression changed between SKCM samples and these samples as GBM, we performed a differential expression analysis using \texttt{edgeR} \cite{Robinson2010-xv, McCarthy2012-ge}. This is an R package that uses a negative binomial distribution model to analyze significant gene expression changes between two groups \cite{Robinson2010-xv, McCarthy2012-ge}. Although normally \texttt{edgeR} works with raw read counts, more recently the package creator has stated that RSEM normalized reads are also suitable for use with \texttt{edgeR} \cite{Smyth2015-ei}. 

We applied the inverse transformation of our min-max normalization to our four interpolated datasets since our UFDN decodes gene expression levels to the range [0,1]. Then we used \texttt{edgeR} to find differentially expressed genes between SKCM samples and 100\% GBM interpolated samples. A p-value threshold for differential expression was set at $p=.05/20501=2.438*10^{-6}$ to control for false discovery.

Analyzing every single gene the significantly changed between SKCM and GBM would be a challenge, so we used integrative Non-negative Matrix Factorization (IntNMF) to learn metagenes that summarized gene expression changes \cite{Chalise2017-zl}. IntNMF learns a reduced dimensionality representation across multiple datasets \cite{Chalise2017-zl}. IntNMF learns a shared basis matrix $W\in\mathbb{R}^{p\times k}$ and  where $p$ is the number of features (here, the differentially expressed genes) and $k$ is the number of metagenes, $k<<p$. Each dataset $D_j$ is described by a learned matrix $H_j\in\mathbb{R}^{k \times n}$ where $n$ is the number of samples in the dataset \cite{Chalise2017-zl}. Each row of $H_j$ represents the linear combination of metagenes of $W$ that combine to reconstruct the original sample in $D_j$ \cite{Chalise2017-zl}. We chose $k=60$ based on an analysis of the reconstruction error $\sum_j ||D_j-WH_j||_F$, where $F$ is the Frobenius norm. We learned $W$ and $H_j$ for each dataset using the R package \texttt{IntNMF} \cite{Chalise2017-zl}.

Every element $g$ of column $W^{(i)}$ is non-negative and represents the contribution of gene $g$ to the $i$-th metagene \cite{Chalise2017-zl}. Each element $s$ of the $n$-th row of $H_j$ represents the contribution of metagene $s$ to the $n$-th sample of the $j$-th dataset. We can analyze how these metagenes change over the different interpolation datasets in order to understand how gene expression is changing \cite{Chalise2017-zl}. 

Finally, to understand the broad composition of the metagenes discovered by IntNMF, we used Gene Ontology (GO) enrichment analysis. GO terms are an ontology of three categories: biological processes, molecular function, and cellular component. They link together information about the functions and relationships of genes and proteins. \texttt{topGO} is an R package that analyzes if GO terms, which have been mapped to genes, show up more often than expected in a set of genes and associated scores for each gene \cite{Alexa2010-od}. 

We used a Kolmogorov-Smirnov like test known as Gene Score Enrichment Analysis that calculates p-values of enrichment based on a score for each gene\cite{Alexa2010-od}. In our work, we did this test on each metagene derived from IntNMF with the score for gene $g$ as $W^{(i)}_g$ \cite{Alexa2010-od}. By looking at the top scoring GO terms for each metagene, we understand what sort of genes are changing as we interpolate between cancer types \cite{Alexa2010-od}. 

\subsection{Code}
All our code is available at \url{https://github.com/bkompa/UFDN-TCGA}. We used Liu et al.'s implementation of UFDN as a starting point but had to expand the architecture to work with an arbitrary number of domains. We wrote all other code used for analysis with the various packages mentioned above. 

\section{Results}
\subsection{UFDN Training and Performance}
\begin{figure}[htbp]
    \centering
    \includegraphics[width=.8\textwidth]{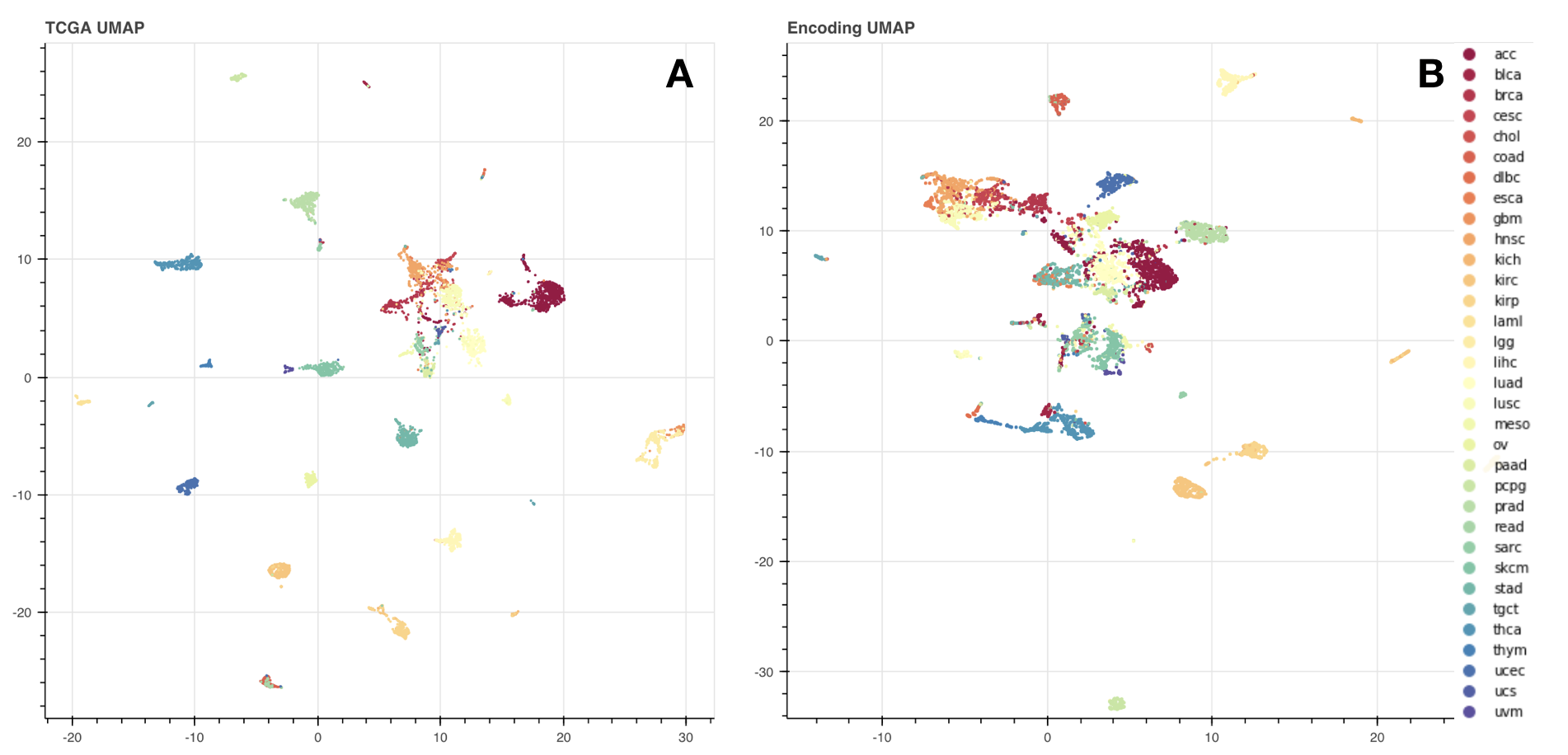}
    \caption{UMAP projections of the RNA-Seq TCGA data (Figure \ref{fig:umap}A) and UFDN latent space encodings of said data (Figure \ref{fig:umap}B). The full 20,501 dimensional representation of gene expression levels have more cancer specific clusters. The 100 dimensional latent space encodings of these samples still clustered in the UMAP space, though to a lesser extent.}
    \label{fig:umap}
\end{figure}

First, we validated that our UFDN learned a non-trivial latent space representation of TCGA RNA-Seq data. We projected both the TCGA data and latent space encodings into UMAP space \cite{McInnes2018-sz}. UMAP learns a Riemann manifold representation of the data \cite{McInnes2018-sz}. We used hyper-parameters \texttt{spread=2.0} and \texttt{min\_dist=.01} to produce Figure \ref{fig:umap}. We observed distinct clusters by cancer types for both the original data and encodings. We proceeded in our downstream analysis confident that our UFDN had learned how to discern between cancer types based on these UMAP projections. 

\begin{figure}[htbp]
    \centering
    \includegraphics[width=.8\textwidth]{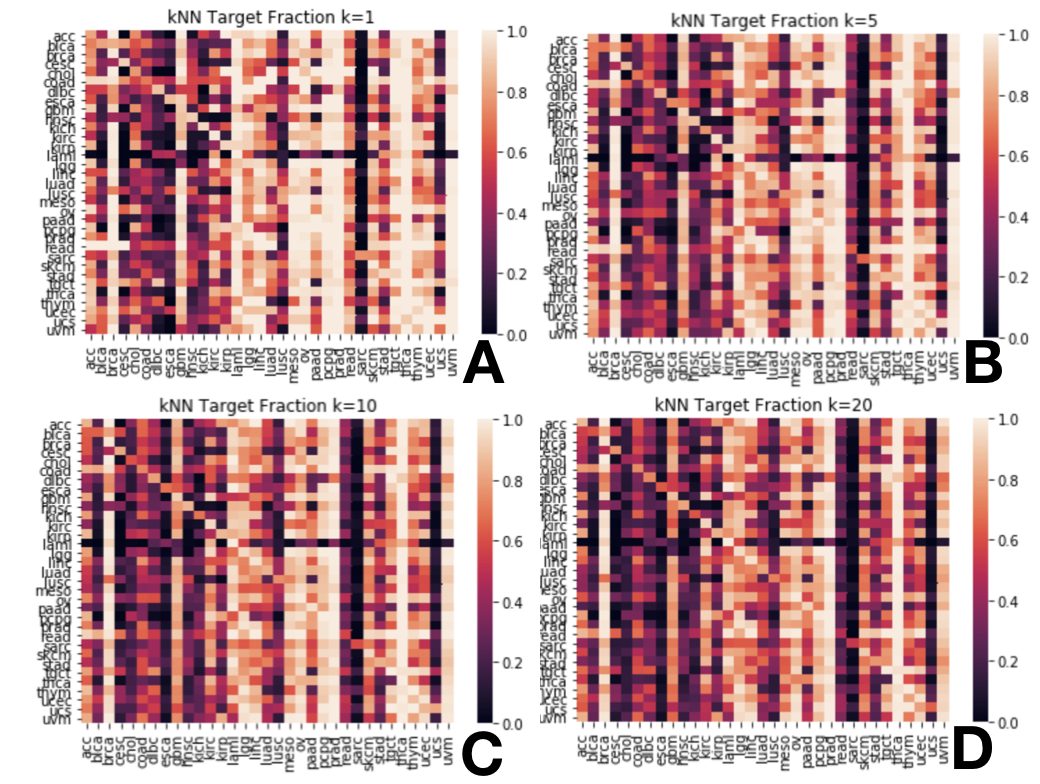}
    \caption{The fraction of $k$ nearest neighbors that were in the target domain (the rows of the figures) after decoding from a source domain (the columns of the figures). Some domains were noticeably more difficult to interpolate into. Glioblastoma had strong interpolation results across $k\in[1,5,10,20]$.}
    \label{fig:kNN}
\end{figure}

Next, we estimated the ability of our UFDN to take data from a source domain (original cancer type) and interpolate these data into a target domain (new cancer type). We considered the fraction of the $k$ nearest neighbors, in the training data, of the interpolated samples that were in the target domain as a measure of success. These decoding rates are shown in Figure \ref{fig:kNN}. There were certain cancers that the UFDN was able to more robustly interpolate into. These included glioblastoma, acute myloid leukemia, mesothelioma, and prostate adenocarcinoma, among others. Difficult cancers to interpolate into were sarcomas, which are a heterogeneous subcategory of soft tissue cancers and cervical squamous cell carcinoma. 
%

\begin{table}[htbp]
\centering
\caption{Results on two classification tasks compared to a random forest baseline.}
\begin{tabular}{@{}llll@{}}
\toprule
Algorithm            & Cancer Status Acc (Train/Test) & Cancer Type Acc (Train/Test) &  \\ \midrule
Random Forests       & \textbf{99.60\%/98.41\%}       & \textbf{99.65\%/95.20\%}     &  \\
UFDN-MSE             &         ---                       &             96.51\%/94.10\%                 &  \\
Unsupervised UDFN    &        95.60\%/86.14\%                          &     ---        &  \\
Semi-supervised UDFN & \textbf{99.60\%/98.41\%}       & 96.51\%/94.10\%              &  \\ \bottomrule
\end{tabular}

\label{tbl:classification}
\end{table}

Finally, we analyzed our UFDN's performance on two classification tasks: cancer status prediction and cancer type prediction. Table \ref{tbl:classification} reports the performances of our three UFDN classification algorithms as compared to a random forest baseline. The random forests had a maximum depth of 15 and were composed of 100 trees. The semi-supervised UFDN algorithm was able to match the performance of random forests on the cancer status task and was comparable on the cancer type task. Other UFDN algorithms were less successful compared to the baseline. 

\subsection{Gene Expression Changes}
\begin{figure}[htbp]
    \centering
    \includegraphics[width=.7\textwidth]{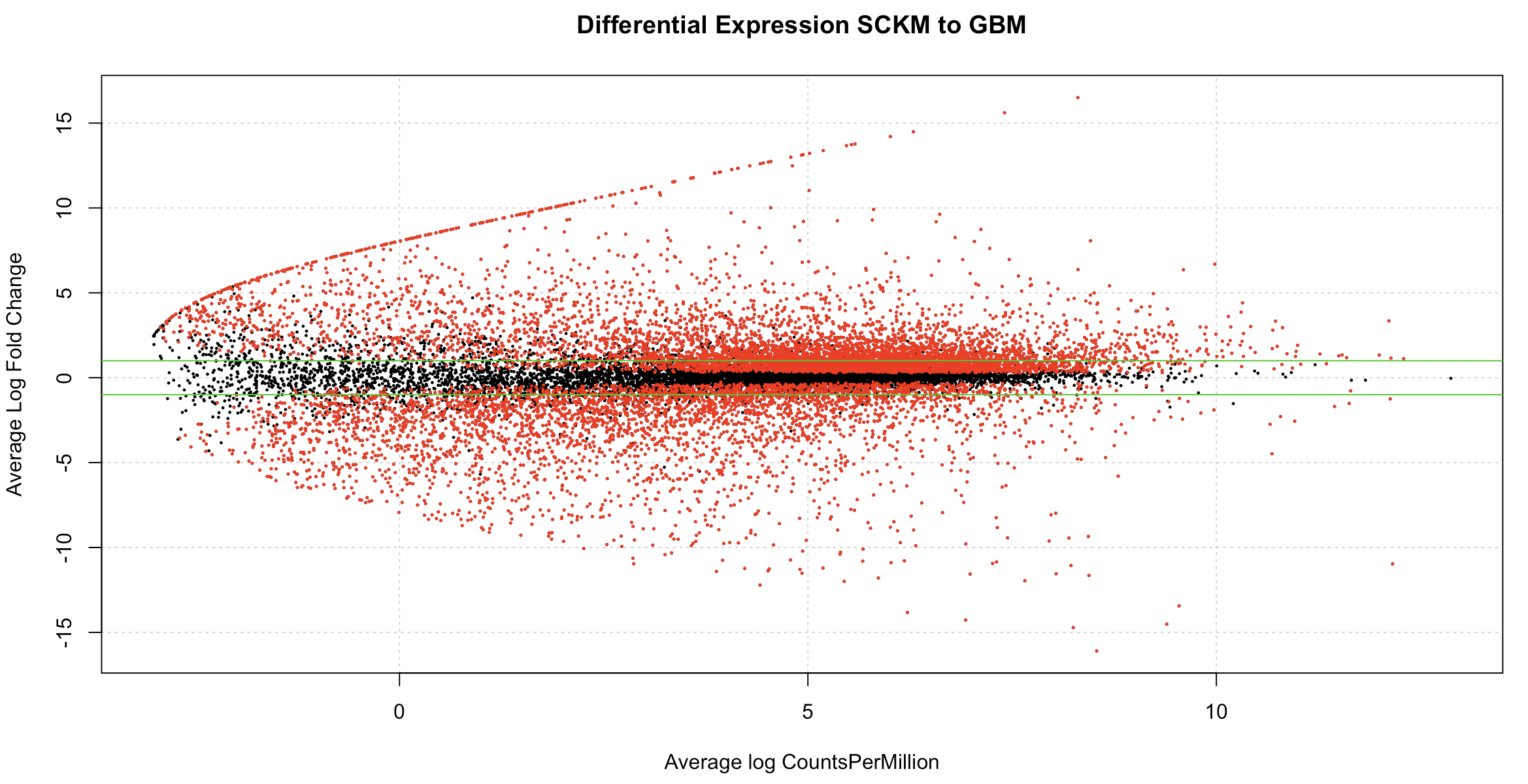}
    \caption{Differential expressed genes at a Bonferroni corrected p-value of $2.43*10^{-6}$. These genes are shown in red, while non DE genes are in black. 10,557 genes were differential expressed between skin cancer samples and 1.0-glioblastoma interpolated skin cancer samples.}
    \label{fig:edgeR}
\end{figure}

After interpolating 95 samples of SKCM from the test set into GBM, we analyzed which genes had significant changes in expression between the SKCM and 1.0-GBM samples. Using \texttt{edgeR}, we looked for genes that had differential expression that exceeded a significance threshold of $p=2.43*10^{-6}$. There were 10,557 genes that exceeded this threshold. Figure \ref{fig:edgeR} shows the plot of average log fold change versus average log counts per million and highlights the differentially expressed genes between the two groups. 

\begin{figure}[htbp]
    \centering
    \includegraphics[width=.8\textwidth]{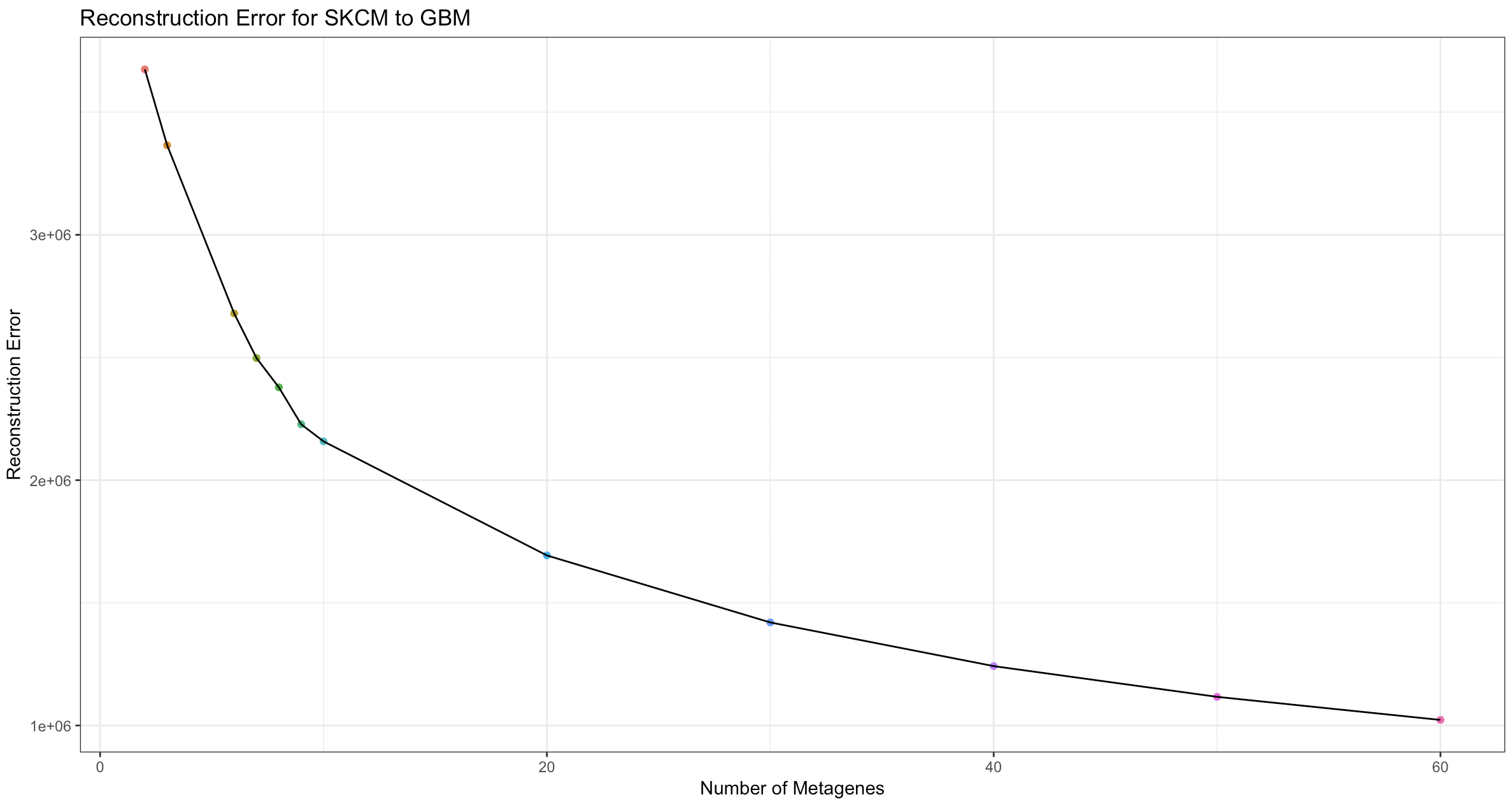}
    \caption{Reconstruction error based on the Frobenius norm from IntNMF versus $k$, the rank of $W$ and $H_j$ in IntNMF, on the x-axis as the number of metagenes. For subsequent analysis, $k$ was chosen to be 60 as error is nearly at an inflection point and plateauing.}
    \label{fig:reconstruction}
\end{figure}

For the 10,557 differential expressed genes, we learned a shared basis $W$ using IntNMF. By varying the rank of that basis, we were able to decrease the reconstruction error across datasets SKCM, 0.25-GBM, 0.5-GBM, 0.75-GBM, and 1.0-GBM. Figure \ref{fig:reconstruction} reports how $k$ affected the reconstruction error. We chose $k=60$ for subsequent analysis based on the inflection point of this reconstruction curve. Hutchins et al. suggest that this is an optimal way to select $k$ for NMF \cite{Hutchins2008-oc}. $k=60$ was also chosen for computational considerations. Optimizing $H_j$ and $W$ for $k=60$ took nearly 7 hours and increasing $k$ much more would significantly increase this considerable time requirement.  

\begin{figure}[H]
    \centering
    \includegraphics[width=\textwidth]{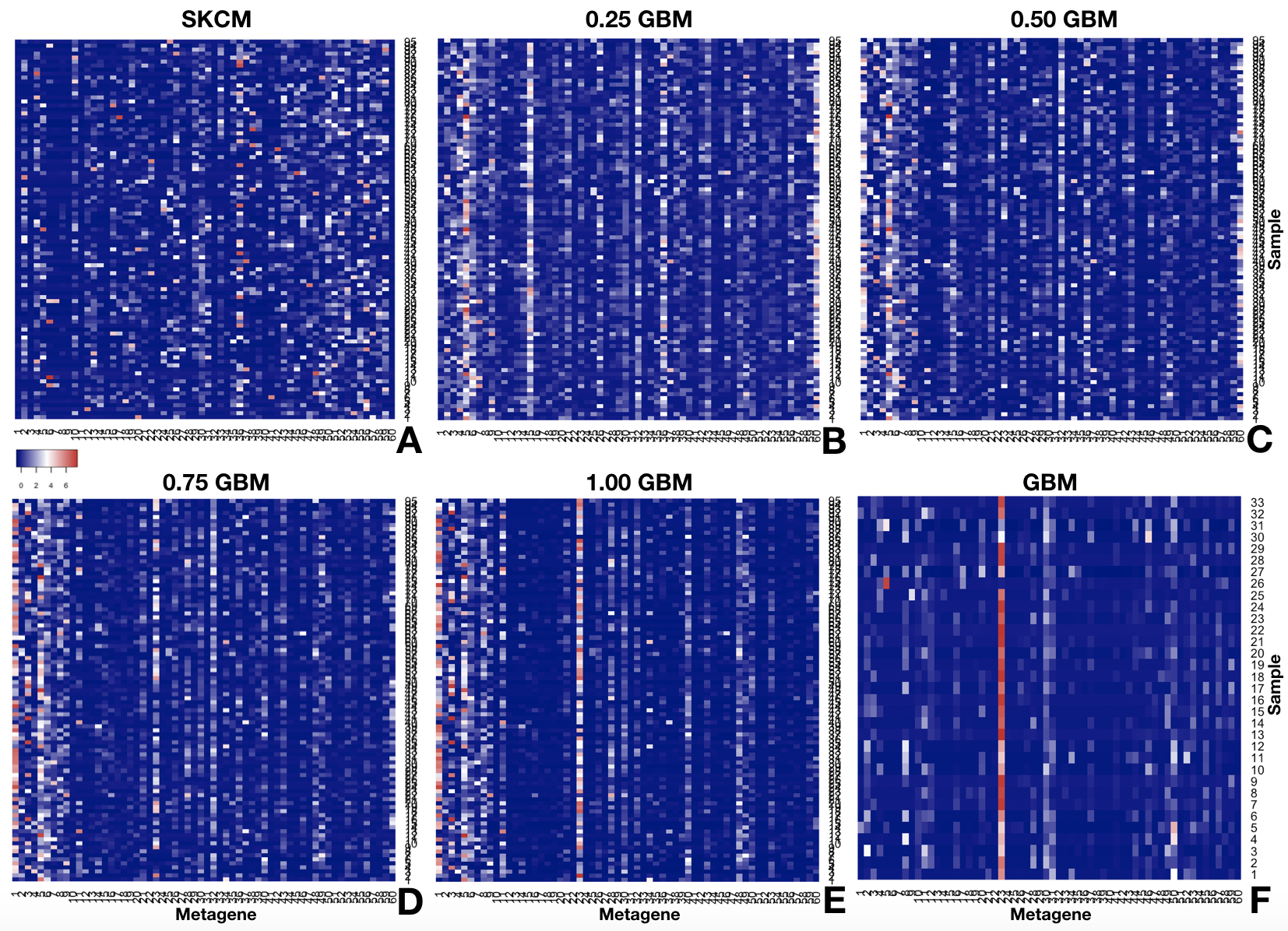}
    \caption{Heatmap visualization of the $H_j$ matrices for each interpolation of the SKCM test data set. No row or column reordering was done to keep consistent metagene order across datasets. A full interpolation of SKCM data into GBM data results in a consistent activation of metagene 23 (Figure \ref{fig:sample_heatmap}E). This is replicated in $H_{GBM}$ (Figure \ref{fig:sample_heatmap}F), which was optimized against the fixed $W$ basis learned for the other 5 datasets.}
    \label{fig:sample_heatmap}
\end{figure}

Finally, we visualized the rows of $H_j$ for each dataset in $\{$SKCM, 0.25-GBM, 0.50-GBM, 0.75-GBM, 1.00-GBM$\}$. The columns of each heatmap in Figure \ref{fig:sample_heatmap} represent the relative activation of the respective metagene. As interpolation towards GBM increases, distinct metagenes increase their responsibility for reconstructing $H_j$. In SKCM, metagene 36 has the most representation in the data. For 0.25-GBM, 0.50-GBM, and 0.75-GBM, it was metagenes 15, 32, and 1, respectively.

In the 1.00-GBM heatmap (Figure \ref{fig:sample_heatmap} E), we saw the increased activation of metagene 23. When we took 33 samples of TCGA GBM data from the test set and learned the matrix $H_{GBM}$ that minimized reconstruction error $||D_{GBM}-WH_{GBM}||_F$ for the same, fixed, $W$ learned previously by IntNMF, we observed the same metagene 23 dominating (Figure \ref{fig:sample_heatmap} F). 

We proceeded to analyze the dominant metagene for every dataset $H_j$ for GO term enrichment. In the interest of space, we only report the top 15 most enriched GO terms for metagene 23 based on p-value. Table \ref{Table2} reports the GO term as well as p-value for each term.

\begin{table}[htbp]
\centering
\caption{The top 15 Gene Ontology Terms enriched in metagene 23}
\begin{tabular}{@{}lll@{}}
\toprule
GO ID      & Term                                        & p-value  \\ \midrule
GO:0003676 & Nucleic acid binding                        & 5.20E-19 \\
GO:0003735 & Structural constituent of ribosome          & 2.70E-15 \\
GO:0003723 & RNA binding                                 & 3.90E-14 \\
GO:0003677 & DNA binding                                 & 1.60E-12 \\
GO:0005198 & Structural molecule activity                & 3.80E-12 \\
GO:0000981 & DNA-binding transcription factor activit... & 4.70E-12 \\
GO:0003700 & DNA-binding transcription factor activit... & 3.50E-11 \\
GO:0140110 & Transcription regulator activity            & 2.80E-09 \\
GO:0008376 & Acetylgalactosaminyltransferase activity    & 4.10E-08 \\
GO:0043492 & ATPase activity, coupled to movement of ... & 1.00E-07 \\
GO:0060089 & Molecular transducer activity               & 1.30E-07 \\
GO:0004126 & Cytidine deaminase activity                 & 2.10E-07 \\
GO:0019239 & Deaminase activity                          & 4.50E-07 \\
GO:0048020 & CCR chemokine receptor binding              & 7.30E-07 \\
GO:0008009 & Chemokine activity                          & 8.10E-07 \\ \bottomrule
\end{tabular}
\label{Table2}
\end{table}
\section{Discussion}
Our UFDN was able to learn a biologically relevant latent space encoding of TCGA data. Classification task results in Table \ref{tbl:classification} indicate that our UFDN was able to compete with random forests that were trained on all 20,501 gene expression features. This indicates our algorithm was able to learn an efficient, useful embedding of gene expression data. Figure \ref{fig:umap} demonstrates that we learned an encoding space that clustered cancers of the same domain. This likely facilitates successful interpolation and classification between cancer domains. Additionally, our UFDN could robustly interpolate into many cancer domains. Although Figure \ref{fig:kNN} demonstrates that not every cancer domain was easy for the UFDN to decode into, one thing to note is that almost every column (target domain) had at least one element with high decoding fraction. It's possible to consider multiple interpolations and potentially from converting from domain A to B to C would have a higher success rate than going from A to C. 

We learned 10,557 differentially expressed genes between SKCM and 1.0-GBM interpolated samples as demonstrated in Figure \ref{fig:edgeR}. This reduction in dimensionality allowed us to make IntNMF computationally tractable. The lower number of genes considered in IntNMF, the faster the learning of the shared basis $W$ and dataset specific $H_j$. Analysis of the reconstruction error from IntNMF informed our choice of 60 metagenes (see Figure \ref{fig:reconstruction}). In Figure \ref{fig:sample_heatmap}, we investigated how linear combinations of these distinct metagenes reconstructed samples from many partially interpolated datasets. We observed unique metagenes increasing activation for each partial interpolation. This is an approximation of how gene expression profiles change during metastasis. 

When we learned $H_{GBM}$, the representation of TCGA GBM samples with respect to the basis $W$, something remarkable happened. Note that $W$ was not informed by the TCGA dataset $GBM$ at all. $W$ was simply the shared basis trained by IntNMF on interpolation datasets SKCM (equivalently, 0.00-GBM), 0.25-GBM, 0.5-GBM, 0.75-GBM, and 1.0-GBM.
Yet when $H_{1.0-GBM}$ and $H_{GBM}$ were compared side by side in Figure \ref{fig:sample_heatmap} E\&F, their metagene activation profiles were dominated by the same metagene 23. Therefore, our interpolation from SKCM to GBM successfully recapitulated observed gene expression activity. 

Furthermore, when we explored several of the GO terms identified by a GO term enrichment analysis, metagene 23 was enriched for terms related to glioblastoma. GO:0008376 represents a glycoprotein with a known association to glioblastoma \cite{Zhang2003-vf,Kroes2007-an}. GO:0004126 refers to cytidine deaminase activity. Cytidine deaminase gene therapy has been identified as a potential treatment for glioblastoma\cite{Fischer2005-zd, Miller2002-fa}. GO:0048020 and GO:0008009 are associated with chemokines, which are implicated in glioblastoma development \cite{Zhou2002-qk, Rempel2000-nr}. Our metagenes learned glioblastoma-specific genes and our UFDN interpolated skin cancer samples to glioblastoma. Further analysis of the metagenes activated during interpolations 0.25-GBM, 0.50-GBM, and 0.75-GBM could provide starting points for the investigation of the metastasis pathway from SKCM to GBM. This could help explain the association between melanoma and glioblastoma that is not currently understood \cite{Desai2008-cj, Scarbrough2014-wj}.

\section{Conclusion}
Our UFDN learned a biologically relevant latent space that facilitated meaningful interpolations between cancer domains. Our latent space can be used to generate more examples of transitions between cancers types. Our interpolations from SKCM to GBM have feasible biological interpretations and suggest possible gene expression changes during the mysterious transition from melanoma to glioblastoma. 

\section*{Acknowledgements}
We acknowledge the helpful suggestions of Dr. Devavrat Shah and Flora Meng on this project. Kompa is also indebted to the feedback of Scott Nanda and Kathryn Almon. 

\bibliographystyle{unsrt}
\nocite{}
\bibliography{6867bib}
\end{document}